\documentclass{aa}  

\usepackage{graphicx}
\usepackage{txfonts}
\usepackage{subfigure}
\usepackage{multirow}
\usepackage{cancel}

\usepackage{array} 

\RequirePackage[colorlinks,citecolor=blue,urlcolor=blue,linkcolor=blue]{hyperref}

\usepackage{xcolor}

\usepackage[normalem]{ulem}

\definecolor{bubbles}{rgb}{0.91, 1.0, 1.0}
\definecolor{aquamarine}{rgb}{0.5, 1.0, 0.83}
\definecolor{bubblegum}{rgb}{0.99, 0.76, 0.8}
\definecolor{bluebell}{rgb}{0.74, 0.74, 0.92}
\definecolor{dollarbill}{rgb}{0.72, 0.93, 0.6}

\newcommand{\vx}{\mathbf x}
\newcommand{\vk}{\mathbf k}
\newcommand{\dd}{{\rm d}}
\newcommand{\ii}{{\rm i}}
\newcommand{\mC}{{\mathcal{C}}}
\newcommand{\eps}{{\epsilon}}
\newcommand{\abs}[1]{{\vert #1\vert}}
\newcommand{\PMF}{{P^{\rm MF}}}
\newcommand{\chiMF}{{\chi^{\rm MF}}}
\newcommand{\WMF}{{W_1^{\rm MF}}}

\newcounter{Xtophecounter}
\newcounter{FFcounter}
\newcounter{SAcounter}
\newcounter{RBcounter}

\begin{document}

   \title{The Topology of Rayleigh-Levy Flights in Two Dimensions}

   \author{Reginald Christian Bernardo
          \inst{1,2}\fnmsep
          \and
          Stephen Appleby \inst{2,3}\fnmsep
          \and
          Francis Bernardeau \inst{4,5}\fnmsep
          \and
          Christophe Pichon \inst{4,5,6}\fnmsep\thanks{Corresponding author: pichon@iap.fr}
          }

   \institute{Max Planck Institute for Gravitational Physics (Albert Einstein Institute), Hannover 30167, Germany
    \and Asia Pacific Center for Theoretical Physics, Pohang 37673, Republic of Korea 
         \and
         Department of Physics, POSTECH, Pohang 37673, Republic of Korea
         \and
         Université Paris-Saclay, CNRS, CEA, Institut de physique théorique, 91191, Gif-sur-Yvette, France
         \and
         Institut d'Astrophysique de Paris, 98 bis Boulevard Arago, F-75014 Paris, France
         \and
         Kyung Hee University, Dept. of Astronomy \& Space Science, Yongin-shi, Gyeonggi-do 17104, Republic of Korea \\
             }


 
  \abstract
    {Rayleigh-L\'evy flights are simplified cosmological tools which
 capture {certain} essential statistical properties of the cosmic density field, including hierarchical structures in higher-order correlations, making them a valuable reference for studying the highly non-linear regime of structure formation.
 Unlike standard Markovian processes, they exhibit long-range correlations at all orders.
Following on  recent work on one dimensional flights, this study explores the one-point statistics and {Minkowski functionals (density PDF, perimeter, Euler characteristic)} of Rayleigh-L\'evy flights in two dimensions.
We derive the 
{Euler characteristic in the mean field approximation and the density PDF and iso-field perimeter $W_{1}$ 
in  beyond mean field calculations}, and validate the {results} against  simulations. 
The match is {excellent} throughout, even for fields with large variances, in particular when finite volume effects in the simulations are 
taken into account and when the calculation is extended beyond the mean field.}


   \maketitle
%

\section{Introduction}
 \label{sec:introduction}

The geometry and structure of cosmic fields evolve from initial homogeneity due to expansion and gravitational forces \citep{1986ApJ...304...15B,Review2002}. These fields reflect the expansion history of the Universe and the growth of substructures. Although small-scale structures are influenced by gravitational and baryonic interactions, large-scale structures result primarily from gravitational instabilities, forming the so-called cosmic web made of voids, walls, filaments, and halos \citep[][]{bondetal1996}.

Traditionally, these structures are studied using correlation functions or power spectra \citep[e.g.][]{Peebles+1975,Fry1985}, which provide limited insight into the intricate web-like patterns. Alternative approaches focus on topology and geometry, employing tools such as Minkowski functionals, Betti numbers, and persistent homology \citep[e.g.][]{White1979, 1994ApJ...434L..43M,MeckeBuchertWagner1994,Gay2012,10.1086/152650,Pranav+2017}. Critical points in the density field, where gradients vanish, serve as useful statistical probes and are increasingly studied for cosmological insights, including clustering analyses \citep[e.g.][]{1986ApJ...304...15B,cadiou2020}.

A related yet significant concept is Rayleigh-L\'evy flights \citep{Anomalous2008}, that describe scale-independent non-Gaussian processes and are useful in modeling the motion of dark halos. They differ from standard Markovian processes by exhibiting long-range correlations and hierarchical properties, making them a simplified yet statistically insightful representation of the fully non-linear cosmic density field.  They were first introduced in astrophysics by \cite{Holtsmark1919}   to capture fluctuations in gravitational systems \citep[see][for more recent work]{Litovchenko2021}.
 L\'evy flights are simply defined as a Markov chain point process whose jump probability depends on some power law of the length of the jump alone \citep{1975CRASM.280.1551M,1980lssu.book.....P,1996ApJ...470..131S,2019PAIJ....3...82U}. 
As such, they capture anomalous diffusion and are related to first-passage theory \citep{2010kvsp.book.....K} and Press-Schechter halo formation \citep{10.1086/152650}.

Previous work \cite{Bernardeau:2022sva,Bernardeau:2024ysi} derived from first principle one- and two-point statistics for the critical points of smoothed sampled realisation of these Rayleigh-L\'evy flights in one to three dimensions.
They provided closed-form expressions for Euler counts and their clustering correlations in the mean-field limit, describing their long-separation behaviour. They also presented 
quadratures for calculating critical point number densities.
Their result was validated by Monte Carlo simulations in one dimension only.
Following \cite{Bernardeau:2024ysi}, the present study focuses on validating in two dimensions using the {Minkowski Functionals -- a set of summary statistics comprising the probability density function of the density field, the total perimeter length of iso-field contours, and the Euler characteristic, which together characterize the non-Gaussianity of the field’s one-point distribution.} We restrict ourselves to expansion beyond the mean field for the density PDF and iso-field perimeter, $W_1$, and leave the exploration of quantities such as Euler characteristics or extrema counts in 2D to future work.

The rest of this work proceeds as follows. Section~\ref{sec:model} recalls the main relevant properties of Rayleigh-L\'evy flights. 
Section \ref{sec:2Dlevy} discusses beyond mean field corrections in two dimensions.
Comparison to Monte-Carlo simulation is provided throughout for validation in Section \ref{sec:2Dfields}. 
Section~\ref{conclusion} wraps up. Appendix \ref{sec:bmf} shows the cumulant generating functions beyond the mean field and the corresponding skewness, converging to the exact value. Appendix \ref{sec:app_b} presents a general derivation of the PDF of the iso-field perimeter.
Appendix \ref{sec:numerical_tests} presents tests performed to ensure the robustness of the simulations.

\section{Rayleigh L\'evy flight model} \label{sec:model}

We first briefly review the basic properties of Rayleigh-L\'evy flights and the mean field limit, discussed in detail in \cite{Bernardeau:2022sva,Bernardeau:2024ysi}.

\subsection{Basics}
\label{subsec:basics}

\begin{figure}[h!]
    \centering
    \includegraphics[width=0.85\linewidth]{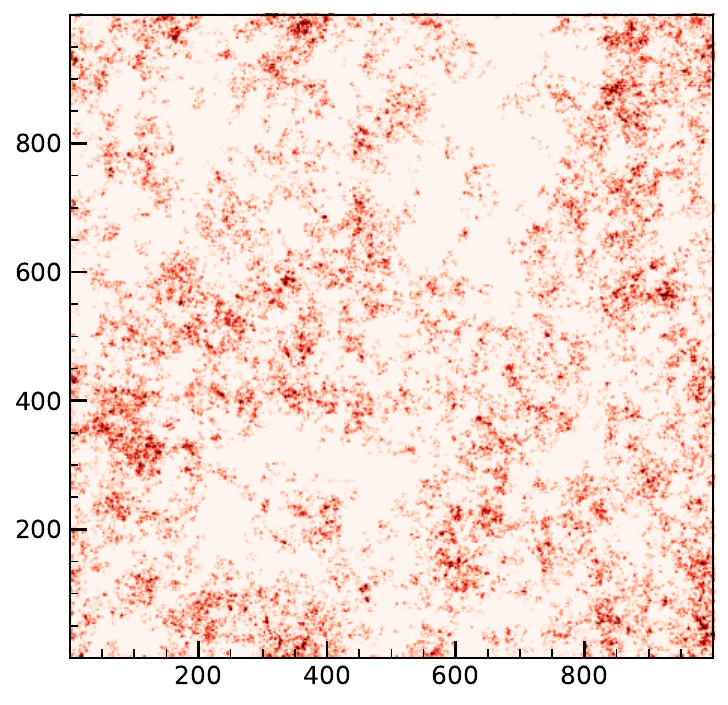}
    \caption{Two-dimensional Rayleigh-L\'evy flight with parameters $\alpha = 3/2$, $\xi_0\sim 2.47$ ($\ell_{0} = 0.01$), density $N_{\rm pts} = 40$ points per pixel, box width $L = 4000 \times \Delta$, resolution $\Delta = 0.25$, and Gaussian smoothing scale $R = 1$.
    }
    \label{fig:Levy_2D}
\end{figure}

{A Rayleigh-L\'evy flight is a point distribution for which each successive point is a step from the previous, in a random direction and with a length drawn from a cumulative probability distribution function of the form}
\begin{equation}
    {\rm CDF}(l) =
    \begin{cases}
    0 \, &, \ \ \ \ l < l_0 \\    
    1 - \left( l_0/l \right)^\alpha \, & , \ \ \ \ l \geq l_0 \,. \end{cases}
\end{equation}
The resulting flights are characterized by rare long jumps, independent of the box size. These are usually handled using periodic boundary conditions. Figure \ref{fig:Levy_2D} illustrates a Rayleigh-L\'evy flight with periodic boundary conditions in two dimensions.

In $D$  dimensions, the two-point correlation functions for flights between positions ${\mathbf r}_{1}$ and ${\mathbf r}_{2}$ is given by
\begin{equation}
\xi({\mathbf r}_{1},{\mathbf r}_{2})
=\frac{1}{n} \int\frac{d^{D}{\mathbf k}}{(2\pi)^{D}}\ e^{i{\mathbf k} \cdot ({\mathbf r}_{2}-{\mathbf r}_{1})}\,\frac{2}{1-\psi(k)}\,,\label{exactxiRL}
\end{equation}
where $n$ is the number density of points and $\psi(k)$ is the Fourier transform of $f_{1}({\mathbf r})$, the density of the subsequent point at position ${\mathbf r}$ (first descendant of a given point at position ${\mathbf{r}_0}$). For a Rayleigh-L\'evy flight, this is given by 
\begin{equation}
f_{1}({\mathbf r})=\frac{\alpha \Gamma(D/2+1)}{2 \pi^{D/2}}\frac{\ell_{0}^{\alpha}}{\vert {\mathbf r} - {\mathbf r}_0 \vert^{D+\alpha}} \,
\end{equation}
for $| {\mathbf r} - {\mathbf r}_0 | > l_0$.
The large distance correlation ($k l_0\ll 1$ or $1-\psi(k) \sim (\ell_0 k)^\alpha$) behaves like
\begin{equation}
\xi({\mathbf r}_{1},{\mathbf r}_{2})
\sim r^{\alpha-D}\ell_0^{-\alpha}/n \,.
\end{equation}
This is subject to constraints ${\alpha < D}$ and ${0 < \alpha < 2}$; the first is required to avoid a logarithmic divergence at ${k=0}$ and the second comes from the expected behavior of $\psi(k)$ at low $k$, ${\psi(k)=1 - (l_0 k)^\alpha + {\cal O}(k^2)}$.\footnote{{This is expected assuming that the correlation function will not exhibit divergent behavior at large distances. The asymptotic behavior can be obtained starting with an ansatz $1-\psi(k) = (l_0 k)^\alpha$ and performing \eqref{exactxiRL} bearing in mind that the correlation function must not diverge at $k = 0$ and large distances $k l_0 \ll 1$.}}

The statistical properties of the field can be inferred through its cumulant generating function (CGF). In principle, having the CGF is tantamount to having the cumulants of the density, its gradient, and the corresponding probability distribution function (PDF), i.e.,
\begin{equation}
\varphi(\lambda_{1},\dots,\lambda_{n})
= {\hskip -0.2cm}
\sum_{p_{1},\dots,p_{n}}\langle\rho_{1}^{p_{1}}\dots\rho_{n}^{p_{n}}\rangle_{c}\,\frac{\lambda_{1}^{p_{1}}}{p_{1}!}\,.
\label{def:CGF}
\end{equation}
In particular, the one-point skewness ${\cal S}_3$ of the density $\rho$ can be derived by expanding the CGF in power of $\lambda\equiv\lambda_\rho$ up to the third order,
\begin{equation}
\label{eq:cgf_cumulants_expansion}
    \varphi(\lambda) = \lambda + \xi_0 \dfrac{\lambda^2}{2!} + {\cal S}_3 \dfrac{\xi_0^2 \lambda^3}{3!} + {\cal O}\left( \lambda^4 \right) \,.
\end{equation}
The PDF of the density can be computed as
\begin{equation}
\begin{split}
    P(\rho) = &\int_{-i\infty}^{+i\infty}\frac{d \lambda}{2\pi i}\exp\left(-\lambda\,\rho+\varphi(\lambda)\right) \\
    = & \int_{-i\infty}^{+i\infty}\frac{d \lambda}{2\pi i} \dfrac{ e^{-\lambda \rho + \varphi(\lambda)} \partial_{\lambda}\varphi(\lambda) }{\rho} \, - \,  { \dfrac{e^{-\lambda \rho + \varphi(\lambda)}}{2\pi i \rho} \bigg |_{-i\infty}^{i\infty} } \,. \notag
\end{split}
\end{equation}
The second line is easier to deal with in practice.

{The remaining piece required to estimate the PDF of the density field is a recipe for the the CGF}. This is provided by the hierarchical tree model (Appendix B of \cite{Bernardeau:2024ysi}). In this case, the CGF can be approximated via the set of equations,
\begin{align}
\varphi(\lambda_{1},\dots,\lambda_{n})
&=\sum_{j}\lambda_{j}
\int{\dd^2{\mathbf x}}\,{\cal C}_{j}({\mathbf x})\left(1+{\tau({\mathbf x})}/{2}\right)\,,
\label{treesumphi}
\end{align}
\noindent and
\begin{equation}
\tau({\mathbf x})=\sum_{j}\lambda_{j}\int{\dd^2{\mathbf x}'}\,{\cal C}_{j}({\mathbf x}')\,\xi({\mathbf x},{\mathbf x}')\,\left(1+{\tau({\mathbf x}')}/{2}\right)\label{treesumtau} \,,
\end{equation}
where $\varphi(\lambda_{1},\dots,\lambda_{n})$ is the CGF, $\tau({\mathbf x})$ is the rooted-CGF (r-CGF), ${\cal C}_j({\mathbf x})$ is a window function in the $j$th cell, and $\xi({\mathbf x}, {\mathbf x}')$ is the two-point correlation function\footnote{We define a `cell' as an area occupied by the smoothing window function, which has typical size $\sim R_{G}$ where $R_{G}$ is the associated smoothing length. We fix $R_{G} = 1$ throughout this work.}. This will be used in the next section to study the properties of Rayleigh-L\'evy flights in the so-called mean field limit.

\subsection{Mean field}
\label{subsec:mean_field}

In the mean field (MF) approximation, the r-CGF is assumed to be constant within each cell. This is reasonable for compact, non-compensated, spherically symmetric density profiles. Denoting by $\tau_i$ the value of the r-CGF for each cell $i$, we average equation~\eqref{treesumtau} over the cell $i$ with profile ${\cal C}_i({\mathbf x})$ to obtain
\begin{equation}
\tau_i\!=\!\sum_{j}\lambda_{j}\,\left(1\!+\!{\tau_j}/{2}\right)\!\int\!{d{\mathbf x}}\!\int\!\!{d{\mathbf x}'}\,{\cal C}_{i}({\mathbf x})\,{\cal C}_{j}({\mathbf x}')\,\xi({\mathbf x},{\mathbf x}')\,.\label{treesumtaudiscrete}
\end{equation}
Then, for the one-point CGF, equation~\eqref{treesumtaudiscrete}
gives the implicit equation
\begin{equation}
\tau_1=\lambda_1\xi_0\left(1-\tau_1/2\right)\,,\label{treesumtaudiscrete1D}
\end{equation}
where $\xi_0$ is the average two-point correlation function in a cell,
\begin{equation}
\xi_0=\int{d{\mathbf x}}\,{d{\mathbf x}'}\,{\cal C}_{1}({\mathbf x})\,{\cal C}_{1}({\mathbf x}')\,\xi({\mathbf x},{\mathbf x}')\,\label{def:xi0}\,.
\end{equation}
The CGF of a Rayleigh-L\'evy flight in the MF limit is therefore
\begin{equation}
\varphi_{1}(\lambda_{\rho};\xi_{0})=\frac{2 \lambda _{\rho }}{2-\xi _0 \lambda _{\rho }}\,.
\label{varphi1}
\end{equation}
Take note that this can be expanded as
\begin{equation}
    \varphi(\lambda_\rho) = \lambda_\rho + \xi_0 \dfrac{\lambda_\rho^2}{2!} + \left( \dfrac{3}{2} \right) \dfrac{\xi_0^2 \lambda_\rho^3}{3!} + {\cal O}\left( \lambda_\rho^4 \right) \,.
\end{equation}
This gives the skewness in the MF approximation to be ${\cal S}_3 = 3/2$. The inverse Laplace transform of \eqref{varphi1} can also be derived analytically, and gives the corresponding PDF of the density
\begin{align}
P(\rho)=e^{-2/\xi _0}\delta^{(D)}(\rho)+\frac{4} {\xi _0^2}e^{-\frac{2 (\rho +1)}{\xi _0}} \, _0{\tilde F}_1\left(2;\frac{4 \rho }{\xi _0^2}\right)\,, \label{eq:P1Dexplicit}
\end{align}
where $\, _0{\tilde F}_1(a;x) = \, _0{F}_1(a;x)/\Gamma(a) = x^{\frac{1-a}{2}} I_{a-1}\left(2  \sqrt{x}\right)$ is the regularized confluent hypergeometric function, $\, _0{F}_1(a;x)$ is the confluent hypergeometric function, $\Gamma(a)$ is the gamma function, and $I_a(x)$ is the modified Bessel function.

The above results are expected to be valid in $D$ dimensions. In \cite{Bernardeau:2024ysi}, the 1D case has been studied extensively with simulations in order to motivate the analytical results. Corrections to the MF were accommodated by a series of approximations, determined by the window function and the two-point correlation function. In Section \ref{sec:2Dlevy} we present analogous beyond mean field (BMF) corrections in 2D.

\section{2D Rayleigh L\'evy flights}
\label{sec:2Dlevy}

\begin{figure}
    \includegraphics[width=0.49\textwidth]{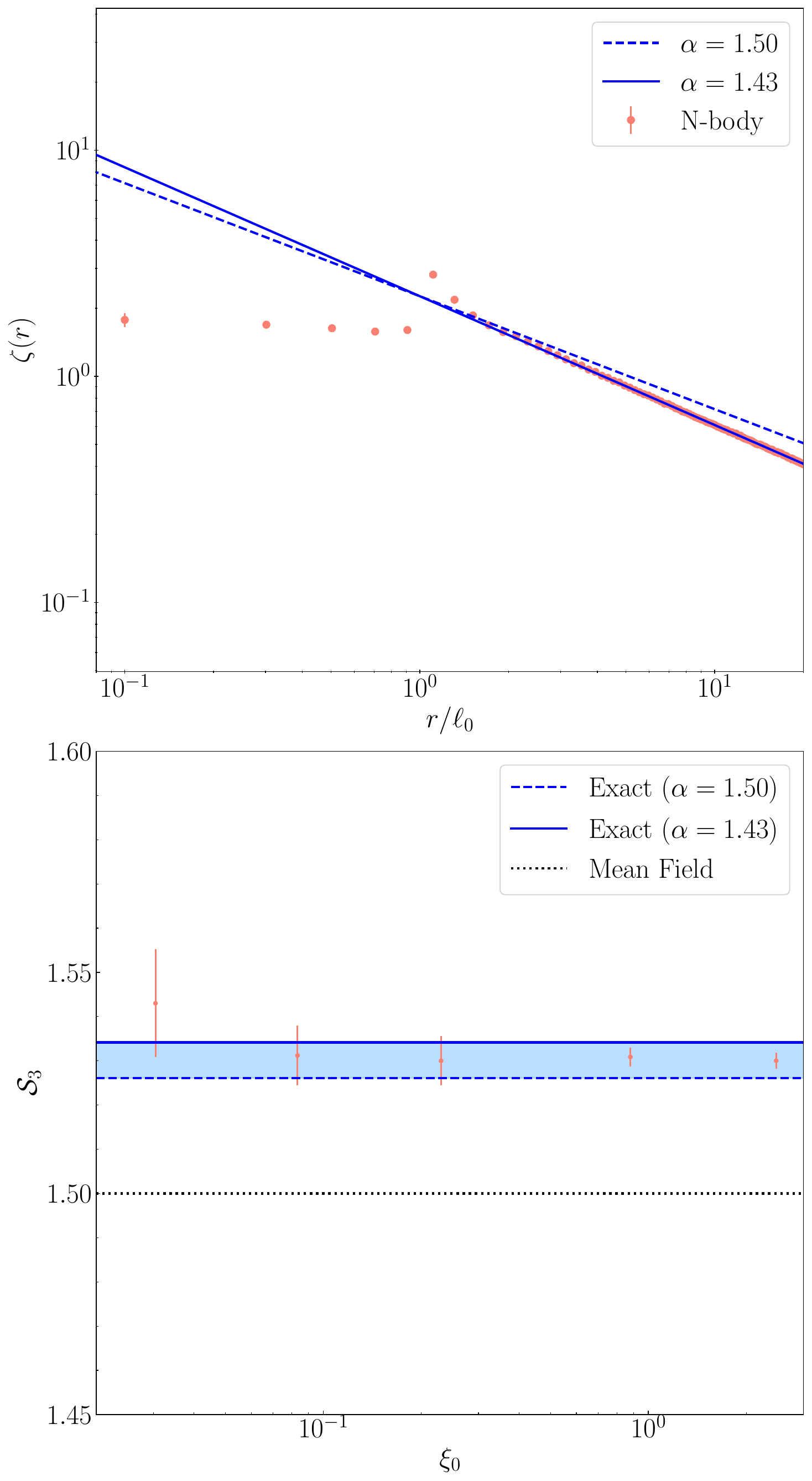}   
    \caption{{\sl Top panel:} measured correlation function for L\'evy flights with $\alpha = 1.5$, $\ell_{0} = 0.02$ (red points/error bars). The blue dashed line is the expected large $r \gg \ell_{0}$ behaviour $\zeta(r) \sim r^{\alpha-D}$, but the actual dependence is slightly steeper, with an effective measured power law $\alpha \simeq 1.43$ (solid blue line). {\sl Bottom panel:} the measured Skewness ${\cal S}_3$ from L\'evy flights with different $\xi_{0}$ variances ({red points and error bars}) with $\alpha = 3/2$. The error bars correspond to the error on the mean from $N_{\rm real} = 100$ realisations each. The beyond mean field prediction yields a better fit to the measured skewness compared to the mean field (black dashed line), but there is a slight ambiguity in recovering the exact BMF prediction due to the $\alpha$ dependence of the statistic (cf. blue solid/dashed lines and shaded region). 
    }
    \label{fig:2D_s3xi0}
\end{figure}

Recall that the MF limit hinges on the assumption that the r-CGF is uniform in a cell, or in symbols, $\tau({\mathbf x})\sim$ Constant. 
However, the MF approximation will not hold exactly. It would be the case if $\xi(\vx,\vx')$ was constant. It is possible build a beyond mean field expansion of the solution of equation \eqref{treesumtau} by lifting 
the constant r-CGF assumption. It requires however to find a proper basis to decompose $\tau({\vx})$ into. One way of doing so is by expanding the r-CGF with the help of a discrete set of function that forms a basis, at least for the type of solutions we are expecting, that is such that $\tau({\vx})\,{\cal C}_{0}({\vx})$ is bounded.

For a Gaussian window function, we have
\begin{equation}
    {\cal C}_{0}({\bf r}) \equiv {\cal C}_{0}(x) {\cal C}_{0}(y) = e^{-r^2/2}/(2\pi) \,,
\end{equation}
where $r^{2} = |{\bf r}|^{2}$. Similarly to the 1D case, the eigenfunctions of the Hamiltonian of a 2D harmonic oscillator can be utilized. In general those solutions will be polynomials in both $x$ and $y$ multiplied by ${\cal C}_{0}({\vx})$. The overall degree of the polynomial gives the total order\footnote{The total order $n_{t}$ corresponds to the energy level $(1+n_{t})E_{0}$ of the associated harmonic oscillator.} $n_{t}$.
Alternatively one can use angular coordinates, $(r,\theta)$, where $x=r\cos(\theta),\ y=r\sin(\theta)$, to describe the solution of a 2D harmonic oscillator. This leads to a choice of basis of the form
\begin{equation}
b_{n}^{\pm m}({\mathbf x})=c_{n}^{m}\,L_{n}^{(m)}(r^{2}/2)\,r^{m}\,e^{{\pm\ii m \theta}}\,,
\end{equation}
where $n$ and $m$ are positive integers, $L_{n}^{(m)}$ are the generalized Laguerre polynomials and the coefficients $c_{n}^{m}$
are normalisation coefficients such that
\begin{equation}
\label{eq:orthodef}
    \int \dd^{2}{\mathbf x}\,
    {b_n^{(m)}({\mathbf x})}\,
    {b_{n'}^{(m')}({\mathbf x})}\,{\cal C}_{0}({\mathbf x})=\delta_{n}^{n'}\delta_{-m}^{m'} \,.
\end{equation}
In this representation the total order is $n_{t}=2 n+m$.
Then the proposition is to expand $\tau(\vx)$ on this basis, taking advantage of the fact that we need to know $\tau(\vx)$ in the vicinity of the origin only, up to scales comparable to the smoothing scales, and that these set of functions form a orthonormal basis
\begin{equation}
\label{eq:r_cgf_expansion}
    \tau({\vx})=\sum_{n,m}
    \tau_n^{m}\,{b_n^{(m)}({\vx})} \,.
\end{equation}
Furthermore the window functions we need to consider can also be built from the same basis. If we set 
\begin{equation}
\label{def:mCnm}
    \mC_{n}^{(m)}({\vx})=L_{n}^{(m)}(r^{2}/2)\,r^{m}\,e^{{\pm\ii m \theta}}\mC_0(r)\,,
\end{equation}
then the components of filtered density gradient are obtained with the help of $\mC_{0}^{(1)}(\vx)$.  
Eq.~\eqref{treesumtau} can then be transformed into a (closed) system in $\tau_n^m$, after 
multiplication by ${\cal C}_0({\mathbf x}) b_{n}^{(m)}
({\mathbf x})$ and integration over ${\mathbf x}$
\begin{equation}
\label{eq:fulldiscretetau}
\tau_n^m=\sum_{n_cm_c}\lambda_{n_cm_c}\left(\Xi_{n0n_c}^{m0m_c}+\sum_{n'm'}\Xi_{nn'n_c}^{mm'm_c}\frac{\tau_{n'}^{m'}}{2}\right)\,,
\end{equation}
where 
\begin{equation}
  \!\!  \Xi_{nn'n_c}^{mm'm_c}\!\!=\!\! 
\int\!\!\dd^2\vx\,\dd^2\vx'\mC_0(\vx)
    {b_n^{(m)}\!({\vx})}\xi(\vx,\vx')
    {b_{n'}^{(m')}\!({\vx'})}
    \mC_{n_c}^{(m_c)}(\vx').\label{eq:defXi}
\end{equation}
\begin{figure*}
    \centering
    \includegraphics[width=0.9\linewidth]{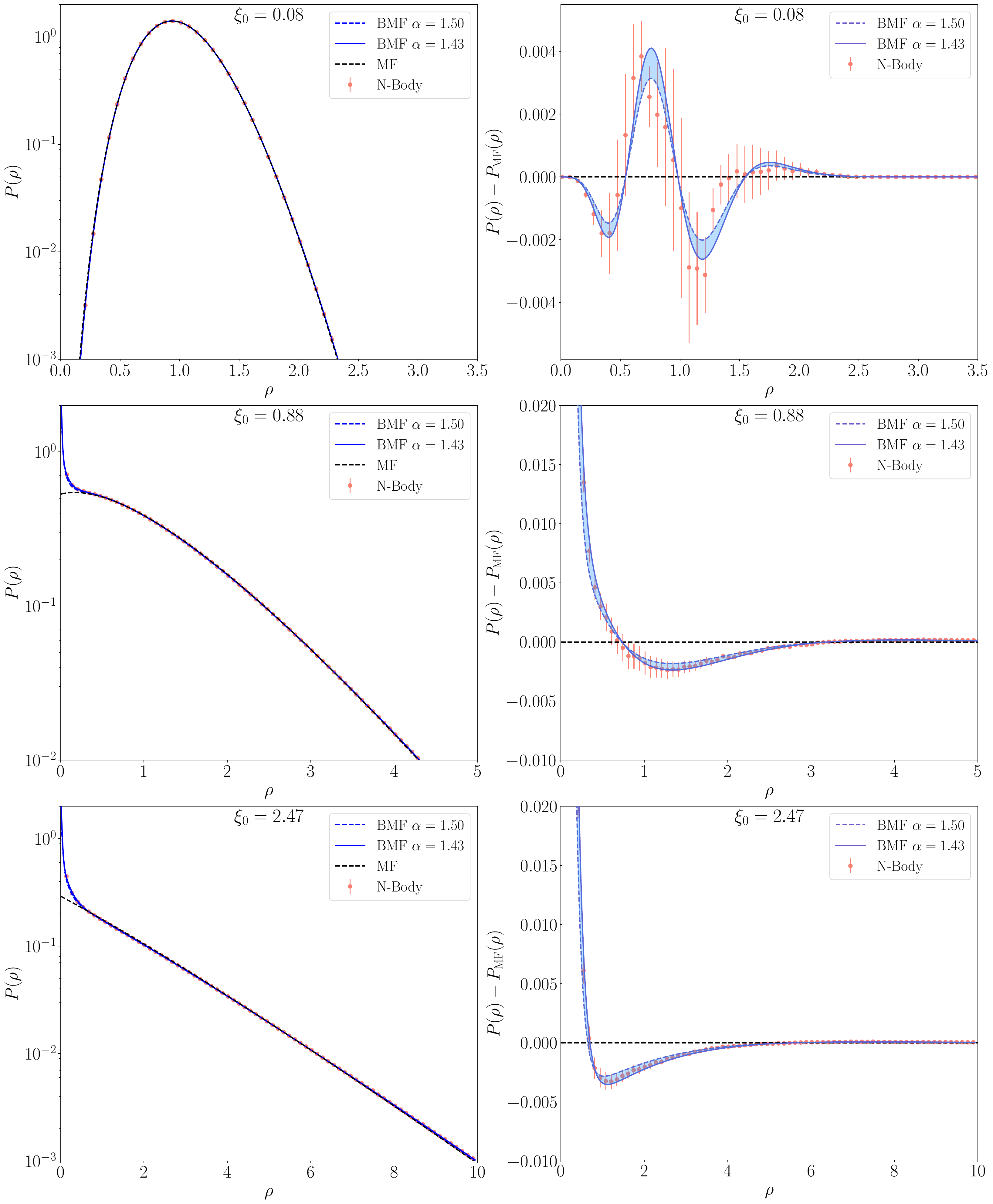}\\
    \caption{{\sl Left:}  Two-dimensional L\'evy flight probability distribution function and {\sl right:} difference with respect to `mean field' (MF) with parameters $\alpha=3/2$, $\xi_0=0.08, 0.88, 2.47$ ({\sl top-bottom panels}), $L = 4000\times \Delta$ (box width), $N_{\rm pts} = 40 \times 2000^2$ (number of steps), $\Delta = 0.25$ (resolution), and $R = 1$ (smoothing scale). The error bars show the standard deviation in the left panels and the error on the mean for $100$ realisations in the right panels. The points the sample means. The corresponding standard deviations of the fields are given by $\sigma_{0} = \sqrt{\xi_{0}} = 0.28, 0.94, 1.57$ ({\sl top-bottom}). The BMF prediction agrees well with the simulations even in the non-perturbative, non-Gaussian regime. 
    }
    \label{fig:2D_ell0}
\end{figure*}

\begin{figure*}
    \centering
    \includegraphics[width=0.9\linewidth]{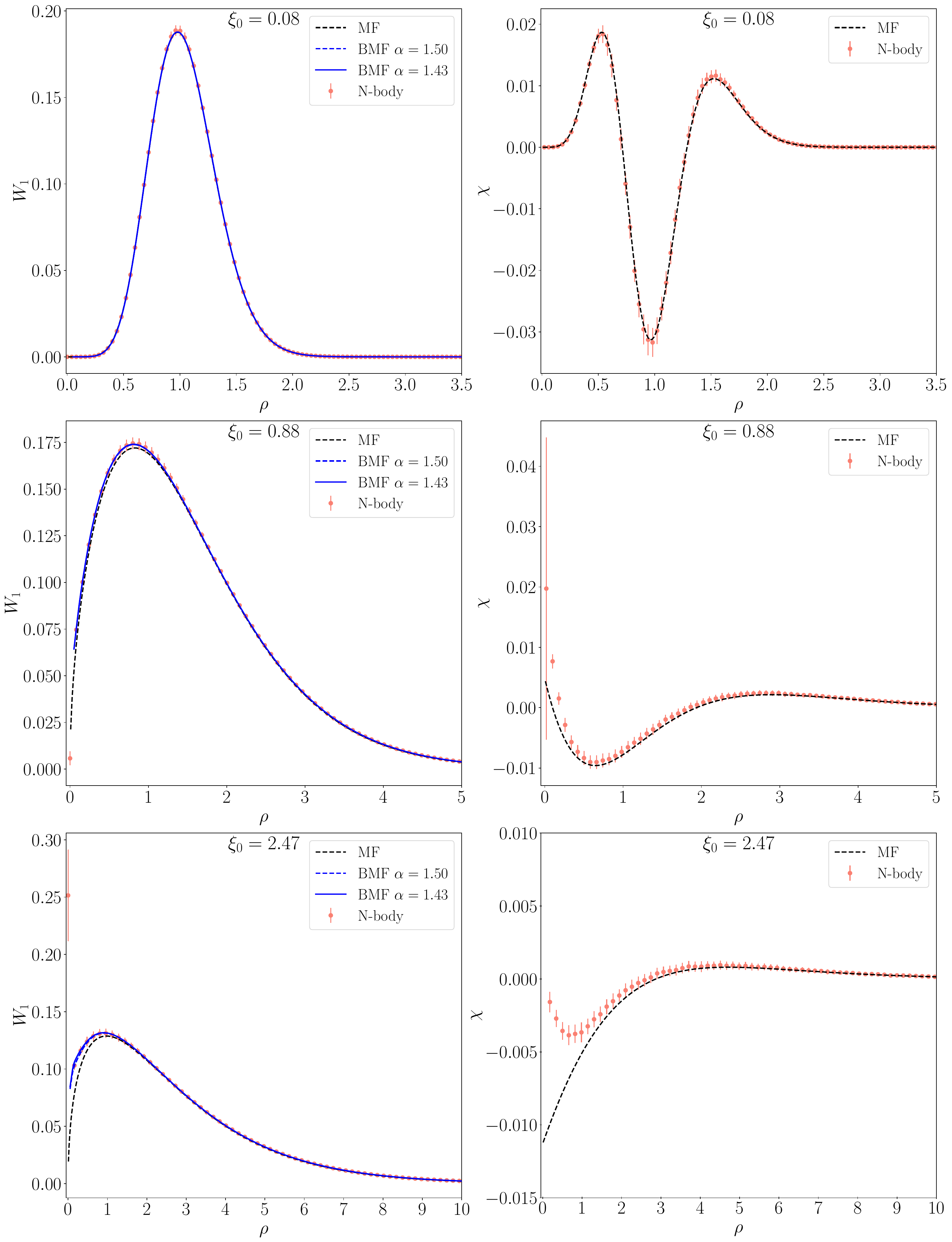}
    \caption{{\sl Left:} excursion set perimeter length $W_{1}$ and {\sl right} Euler characteristic -- measured (red points/error bars), mean-field expectation value (black dashed lines) and beyond mean field with $\alpha =1.5$ (blue dashed lines) and $\alpha = 1.43$ (blue solid lines). The error bars represent the standard deviation, and the points the sample means. {\sl Top to bottom panels} are the cases
    $\xi_0=0.08, 0.88,2.47$,
    respectively, with other parameters $\Delta = 0.25$, $R = 1$, $L=4000 \times \Delta$, $\alpha = 3/2$, $N_{\rm pts} = 40 \times 2000^{2}$. Once again the agreement is excellent, especially at the high density end. Note that the beyond mean field prediction for $\chi$ (right panels) is beyond the scope of this paper.
    }
    \label{fig:MF}
\end{figure*}

Let us  truncate the sum \eqref{eq:r_cgf_expansion} up to $n_t$. 
The algebraic system \eqref{eq:fulldiscretetau}  can then  be solved for $\tau_n^{(m)}$ to obtain the joint CGFs (and, possibly peak statistics, although the latter would generally need to be computed numerically).

The detail of the calculations, involving the final choice of basis and the computation of the required $\Xi_{nn'n_c}^{mm'm_c}$ 
coefficient are given in Appendix \ref{sec:bmf}. We give here  the results obtained in terms of joint CGF. 

For $\alpha = 3/2$, the one-point CGFs of the density field with increasing BMF order can be shown to be
\begin{align}
    \varphi_0(\lambda) &= \frac{2 \lambda }{2-\lambda  \xi _0}\,, \\
    \varphi_2(\lambda) &= -\frac{\lambda  \left(5 \lambda  \xi _0-128\right)}{2 \lambda ^2 \xi _0^2-69 \lambda  \xi _0+128}\,, \\
    \varphi_4(\lambda) & = -\frac{\lambda  \left(225 \lambda ^2 \xi _0^2-59680 \lambda  \xi _0+1048576\right)}{80 \lambda ^3 \xi _0^3-25569 \lambda ^2 \xi _0^2+583968 \lambda  \xi _0-1048576} \, , \notag
\end{align}
{where the index $i$ on $\varphi_i$ indicates the BMF order}. Appendix \ref{sec:cgfs_bmf} spells out higher CGFs, up to $\varphi_6(\lambda)$, together with the skewness ${\cal S}_3$ derived within the BMF formalism up to 3rd order (terms in Eq.~\ref{eq:r_cgf_expansion}) compared with the exact value. 

The CGFs in the BMF formalism can also be obtained for an arbitrary power law correlation function [$\xi(r) = 1/r^\beta$]. The corresponding CGFs can be written analytically as 
\begin{align}
    \varphi_0(\lambda) &= \frac{2 \lambda }{2-\lambda  \xi _0}\,, \\
    \varphi_2(\lambda) &= -\frac{\lambda  (\beta  (\beta +2) \lambda  \xi_0 -32)}{\lambda  \xi_0  (-\beta  (\beta -\lambda  \xi_0 +2)-16)+32} \,.
\end{align}

In the limit $\beta \rightarrow 1/2$, this reduces to the previous expressions with $\alpha=3/2$. The exponent $\beta$ is related to the L\'evy flight index via $\beta=D-\alpha$. For $\beta \rightarrow 0$ (or $\alpha\rightarrow D$), all CGFs at higher order reduce to the mean field value, $\varphi_i(\lambda) \rightarrow \varphi_0(\lambda)=2\lambda/(2-\lambda \xi_0)$ for $i \geq 1$.

Similarly, BMF expressions can be obtained for joint CGF. In practice we derive such expressions to compute the $W_1$ statistics taking advantage of the fact that the latter can be expressed through a two-dimensional numerical integration of the joint CGF for the density and one gradient component. This is in itself a non-trivial result the derivation of which is given in the Appendix \ref{sec:app_b}. There is no such counterpart for the Euler characteristic (to our knowledge). For the latter we could only compare simulations with the MF results.

More precisely, the  BMF for the joint CGF, $\varphi(\lambda,\lambda_x)$,  as a function of variables conjugate to density and $x$-gradient  respectively is given for $n_t=1$ by
\begin{equation}
\varphi_1(\lambda,\lambda_x)\!=\!\frac{64 \lambda \!+\! 8 \beta  \xi _0 \left(\lambda _x^2-\lambda ^2\right)}{\beta  \xi _0^2 \left(4 \lambda ^2\!+\!(\beta\! -\!4) \lambda _x^2\right)\!+\!64\!-\! 8 (\beta +4) \lambda 
   \xi _0}.
   \label{eq:phi1MT}
\end{equation}
Using this expression one can already see departures from the MF predictions. To make the comparisons with numerical results, we use the $n_t=3$ results as shown in the Appendix~\ref{sec:bmf}, Eq~\eqref{eq:phi3}. For the latter a system involving six basis functions is used.

In the next section, we compare the mean field and beyond mean field predictions with simulations through the lens of one-point statistics; {specifically the Minkowski functionals} which are the PDF of the density, the perimeter length $W_1$ and the Euler characteristic $\chi$. The latter two are defined as follows 
\begin{eqnarray} 
\nonumber & & \chi(\rho) \!=\!  \int d\rho_{xx} d\rho_{yy} d\rho_{xy} \left( \rho_{xx}\rho_{yy} \!-\! \rho_{xy}^{2}\right) P(\rho,\rho_{i}\!=\!0,\rho_{ij})\,, \\
 & & W_{1}(\rho) \!=\!  \int d\rho_{x}d\rho_{y} \sqrt{\rho_{x}^{2} + \rho_{y}^{2}} P(\rho,\rho_{i}) \,, \label{eq:defW1}
\end{eqnarray}
where $\rho_{i}$ and $\rho_{ij}$ denote first and second derivatives of the density field. Both admit analytical expressions in the mean field limit. For the Minkowski functional $W_{1}$, this can be derived using the joint PDF of the density and its gradient, \cite{Bernardeau:2022sva},
\begin{equation}
\PMF(\rho,\rho_x,\rho_y) = \PMF(\rho)
\times G\left(\rho_x, \sqrt{\xi_1 \rho}\right) G\left(\rho_y, \sqrt{\xi_1 \rho}\right) \notag
\end{equation}
where $G(\rho, \sigma)$ is the Gaussian PDF with a mean $\rho$ and standard deviation $\sigma$, and $\xi_1$ is the variance  
of the gradient of the density. The expression in the mean field is 
\begin{equation}
\WMF(\rho) = \sqrt{ \pi \xi_1 \rho/2 }\ \PMF(\rho) \,.
\label{eq:W1MF}
\end{equation}

Conversely, the Euler characteristic can be derived in the MF approximation using the joint PDF of the density, the gradient and its second derivatives. The derivation is made  explicitly in Appendix D of \cite{Bernardeau:2024ysi}. This leads to the result
\begin{align}
\chiMF(\rho) &= -\dfrac{\xi_1 e^{-\frac{2 (\rho +1)}{\xi _0}}}{\pi \xi_0^4 \rho} \bigg( \xi_0(\xi_0 + 8\rho) \,_0\tilde{F}_1\left(1, \dfrac{4\rho}{\xi_0^2}\right) \notag \\
& - \left( 2\xi_0 \rho +\xi_0^2 + 8\rho(\rho+1) \right) \,_0\tilde{F}_1\left(2, \dfrac{4\rho}{\xi_0^2}\right) \bigg) \,.\label{eq:chiMF}
\end{align}
For the $W_1(\rho)$ functional it is possible to go beyond the MF approximation and derive general BMF results with the help of a two dimensional integral. The latter involves
$\varphi(\lambda,\lambda_x)$ only (taking  also into account rotational invariance to reduce the dimensionality of the problem) as shown in detail in appendix \ref{sec:app_b}.
The expression we use in practice is the following
\begin{align}
    W_1(\rho)=&-
    \int_{-\ii\infty}^{+\ii\infty} \frac{\dd\lambda_\rho}{2\pi}
    \int_0^\infty\frac{\dd\lambda_g}{\lambda_g}
\frac{\exp\left(-\lambda_\rho\rho\right)}{\rho}\nonumber
\\
&\times\frac{\partial}{\partial\lambda_\rho}\left[
\frac{\partial\varphi(\lambda_\rho,\lambda_g)}{\partial\lambda_g}
\exp\left(
\varphi(\lambda_\rho,\lambda_g)
\right)    \right]\,. \label{eq:W1final}
\end{align}
This equation, together with 
Eqs~\eqref{eq:phi1MT} or \eqref{eq:phi3} (when replacing $\lambda_x$ by
 $\lambda_g$), 
 allows  us to predict $W_1(\rho)$ for the range of values of $\xi_0$ we consider, provided the integrations over $\lambda_\rho$ and $\lambda_g$ are made with sufficient precision.
The Expressions \eqref{eq:W1MF} to \eqref{eq:W1final} are the central theoretical prediction that are tested against  simulations in the following section.

\section{Results}\label{sec:2Dfields}

To validate the results of the previous section, we generate numerical realisations of two-dimensional Rayleigh-L\'evy flights and extract the probability density function of the resulting density fields as well as certain summary statistics; specifically the one-point skewness of the field, the Euler characteristic and the Minkowski Functional $W_{1}$.

The L\'evy flights are generated starting from an initial point in the center of a box of area $L\times L$. The flight is moved a distance $\ell$ from the previous point, in a direction drawn from a uniform angle $0 \leq \theta < 2\pi$. The $i^{\rm th}$ point in the flight is given by $\bf{x}_{i} = {\bf x}_{i-1} +  (\ell_{i} \cos\theta_{i}, \ell_{i} \sin\theta_{i})$. Periodic boundary conditions are enforced, and ${\bf x}_{i}$ is adjusted modulo $L$ if the flight length is greater than $L$. Each point in the flight is binned into a regular lattice using a nearest-neighbour scheme, and we use resolution $\Delta = 0.25$ and number of pixels per side in the box $N_{\rm pix} = 4000$, so that the box size is $L = N_{\rm pix} \times \Delta$. We fix the number of L\'evy flight steps as $N_{\rm pts} = 40 \times 2000^{2}$, $\alpha = 3/2$ and vary $\ell_{0}$ to obtain fields with different $\xi_{0}$ values. Once the L\'evy flight is binned into a regular lattice, we fast Fourier transform the field, apply a Gaussian smoothing kernel $W(kR) = e^{-k^{2}R^{2}/2}$, and inverse transform. We fix $R=1$ to match the 
window function
used in the previous section. {The particular case $\alpha = 3/2$ is studied throughout this work, as it is a numerically tractable example of a Levy flight. Increasing $\alpha$ increases the large scale correlations of the points and generates larger finite box effects, whereas smaller $\alpha$ decreases the typical flight length and requires an increasingly large number of steps to populate the simulations.}

From the Gaussian smoothed L\'evy flight density fields, we generate the one-point skewness, density probability distribution function, the Euler characteristic $\chi$ and iso-density perimeter length $W_{1}$, and compare these statistics to their mean field expectation values and the beyond mean field predictions where available. For details on how we generate these summary statistics from the fields, we direct the reader to Appendix \ref{sec:numerical_tests}. 

Before presenting the results, we first comment on a numerical systematic which hinders our ability to exactly reproduce the theoretical expectation values constructed in the previous section. In Figure \ref{fig:2D_s3xi0} (top panel) we present the two-point correlation function of a set of $N_{\rm real} = 100$ numerically generated L\'evy flights with $\alpha = 3/2$ and $\ell_{0} = 0.02$ (red points/error bars). The expected large separation behaviour of the flight is $\zeta(r) \sim r^{\alpha-D}$ with $D=2$; this expected behaviour is presented as a blue dashed line. However, we find that the simulated L\'evy flights present slightly steeper power law behaviour (solid blue line), corresponding to an effective $\alpha = 1.43$\footnote{{For larger separations, the correlation function drops faster than a power law and becomes increasingly noisy. In Figure \ref{fig:2D_s3xi0} we present only the scales that dominate the statistical properties of the flights, which are well approximated by a power law.}}. This is a finite box effect; each step in the flight must be smaller than $\sim L$ due to the imposition of periodic boundary conditions. This leads to an excess probability of finding flight points at smaller separations, and leads to a systematic difference between theoretical expectation values and our numerical results\footnote{Or alternatively one could model finite volume effects by replacing in  equation \eqref{exactxiRL}  the integral over $\vk$ by a  sum over discrete values.}.
The problem will be exacerbated for larger values of $\alpha$, as the correlation function should become flat as $\alpha \to D$, which is a scenario that cannot be realised for a finite box. In what follows, for all summary statistics we present the expected theoretical results based on $\alpha = 3/2$ and the effective result assuming $\alpha = 1.43$. For mean field predictions, there is no $\alpha$ dependence. 

In Figure \ref{fig:2D_s3xi0} (bottom panel) we present ${\cal S}_{3}$ -- the pixel sample skewness for L\'evy flights with different $\xi_{0}$ variances (red points/error bars). For each $\xi_{0}$ value, we generate $N_{\rm real} = 100$ realisations and present the mean and the error on the mean of ${\cal S}_{3}$. The mean field limit, and exact BMF expectation values are presented as black dotted and blue dashed/solid lines respectively. The solid blue region represents the uncertainty associated with finite box effects. The means of the simulations agree well with the expected exact solutions, to within $\lesssim  0.5\%$ for all variances probed. 

The probability distribution functions $P(\rho)$ are presented in Figure \ref{fig:2D_ell0}. In the left panels we present the numerically reconstructed PDFs (red points/error bars), the mean field expectation value (\ref{eq:P1Dexplicit}) (black dashed lines) and the beyond mean field prediction constructed in Section \ref{sec:2Dlevy} (solid/dashed blue lines), using the third order BMF expansion. In the left panels we present the PDFs, and in the right panels the difference between the numerically reconstructed PDFs and the mean field approximation, and the corresponding difference between BMF and MF theoretical expectation (blue lines). The top/middle/bottom panels correspond to fields with average variance $\xi_{0}=0.08, 0.88, 2.47$ respectively. In the left panels the points and error bars are the mean and the standard deviations of $N_{\rm real}=100$ realisations. In the right panels the error bars correspond to the error on the mean. The solid blue filled regions are the uncertainty associated with finite box effects. We observe that the BMF prediction closely matches the numerical results even for fully non-Gaussian fields with $\xi_{0} > 1$. The mean field approximation fails at low densities, but remains a good fit to the data in the high density tails.

\begin{figure*}
    \centering
    \includegraphics[width=0.9\linewidth]{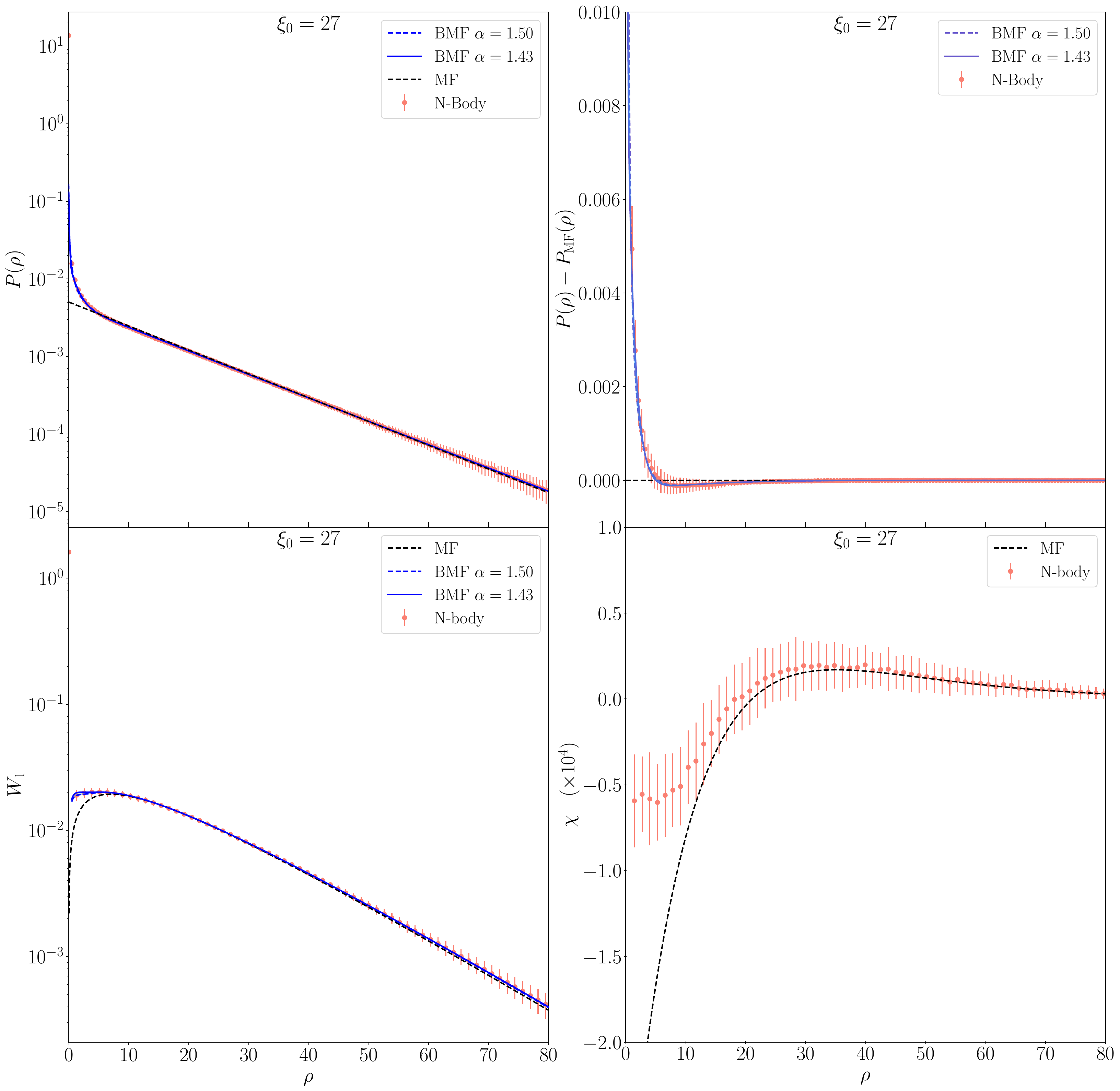}
    \caption{Summary statistics $P(\rho)$ ({\sl top left}), $W_{1}$ ({\sl bottom left}) and $\chi$ ({\sl bottom right}) for a set of $N_{\rm real} = 100$ realisations of a highly non-Gaussian field with $\xi_{0} =27$ and other parameters $\Delta = 0.25$, $R = 1$, $L=4000 \times \Delta$, $\alpha = 3/2$, $N_{\rm pts} = 40 \times 2000^{2}$. The difference between $P(\rho)$ and the mean field prediction is also presented ({\sl top right}). Error bars show the standard deviation, and the points the sample means. The BMF estimates of the PDF (blue solid lines, {\sl top panels}) and $W_{1}$ ({\sl bottom left panel}) remain an excellent approximation for $\xi_{0} > 20$. The mean field limit is a good approximation for all statistics in the high density tails. }
    \label{fig:large_xi0}
\end{figure*}

There are two other Minkowski Functionals associated with a two-dimensional field; iso-field perimeter length $W_{1}$ and Euler characteristic $\chi$. We present these quantities in the left/right panels of Figure \ref{fig:MF} respectively for the same data used in Figure \ref{fig:2D_ell0}. Specifically the top/middle/bottom panels are measurements made from realisations with average field variances $\xi_{0} = 0.08, 0.88, 2.47$ respectively. The red points/error bars are the mean and standard deviations of $N_{\rm real}=100$ L\'evy flight realisations. The mean field predictions are presented as black dashed lines and the BMF predictions for $W_{1}$ are presented as blue-dashed ($\alpha = 1.5$) and blue-solid ($\alpha=1.43$) lines (left panels). The quantities $W_{1}$ and $\chi$ depend on the joint probability distribution function for the field $\rho$ and its first and second derivatives. The joint PDF of the field and its first derivatives, and the corresponding BMF prediction for $W_{1}$, is presented in Section \ref{sec:2Dlevy} and Appendices \ref{sec:bmf},\ref{sec:app_b}. The quantity $W_{1}$ depends on both $\xi_{0}$ and $\xi_{1}$, and we measure the variance of the field $\xi_{0}$ and its gradient $\xi_{1}$ from the simulations and use these values in the analytic form for $W_{1}$. Also, we note that $W_{1}$ is a dimension-full quantity. Throughout this work, we are choosing the smoothing scale $R=1$, so it should be understood that $W_{1}$ is presented in units of $R$. 

In the left panels of Figure \ref{fig:MF} we observe that the mean field prediction for $W_{1}$ (black dashed lines) closely matches the numerical simulations (red points) for small $\xi_{0}$ values (cf. top panel) and the large $\rho$ tails for large variance fields (cf. middle, bottom panels). The BMF prediction agrees well with the simulations for practically all densities, even for field variances larger than unity. The right panels present a similar conclusion for $\chi$; both for small $\xi_{0}$ fields and the high density tails of large $\xi_{0}$ fields, the mean field approximation performs well. The BMF prediction for the joint PDF of the first and second derivatives, required for $\chi$, is left for future work. 

In Figure \ref{fig:large_xi0} we present the same statistics as in Figures \ref{fig:2D_ell0} and \ref{fig:MF}, but now taking a set of $N_{\rm real} = 100$ highly non-Gaussian L\'evy flights with average variance $\xi_{0}=27$. The BMF prediction (blue lines, top and lower left panels) remains an excellent approximation to the numerically reconstructed PDF, and the mean field limit is accurate in the high density tails $\rho \gtrsim 10$. The typical magnitudes of $W_{1}$ and $\chi$ decrease with increasing $\xi_{0}$; this is due to the L\'evy flights occupying an increasingly small number of pixels as we decrease $\ell_{0}$. Most of the pixels are zero density, and as we continue to increase $\xi_{0}$ we will encounter issues with the numerical reconstruction, chiefly with the resolution required to resolve the regions containing L\'evy flight points. For large values of $\xi_{0}$, varying the box size will significantly affect the properties of the field. Decreasing the box size will decrease the number of empty pixels, which will increase the average density and decrease the corresponding variance after $\rho$ is normalised to $\bar{\rho} = 1$. Regardless, our numerical reconstructions confirm that the BMF results are accurate even for highly non-Gaussian flights.

\section{Conclusion} \label{conclusion}

Rayleigh-L\'evy flights belong to a family of point distributions for which the cumulant generating function can be exactly calculated in a certain limit -- specifically the mean field limit -- and as such their statistical properties can be explored in the non-perturbatively non-Gaussian regime. In this work we have measured the statistical properties of two-dimensional Rayleigh-L\'evy flights, to test the underlying accuracy of the mean field and beyond mean field theoretical expectations. For this purpose, we utilized the Minkowski Functionals, which are summary statistics that contain non-Gaussian information to all orders in the one-point statistics of the field. 

Predictions for the PDF of the density field beyond the mean field approximation, obtained by expanding the r-CGF to third order in a Laguerre polynomial basis, were compared to simulations and found to be in agreement at both high and low densities. The PDF was accurately reproduced even for non-perturbatively, non-Gaussian fields with variances larger than unity (Figure~\ref{fig:2D_ell0}). Specifically, we confirm the BMF estimate is accurate over three orders of magnitude for the field variance $10^{-1} \lesssim \xi_{0} \lesssim 10^{2}$. Similarly the BMF estimate for $W_{1}$ was constructed for the first time and shown to be in agreement with simulated Rayleigh-L\'evy flights over the same wide range of field variances. While the BMF prediction for the Minkowski functional $\chi$ is beyond the scope of this paper, the mean-field predictions for this statistic was also favourably compared to Monte Carlo runs. The mean field approximation was found to be a good approximation at intermediate and high densities, even for highly non-Gaussian fields, only failing at low density (Figure~\ref{fig:MF}).  
Our precise match to ensembles of simulations highlighted residual biases,
e.g. associated with finite box effects leading to effective values for $\alpha$ as measured via the correlation function which differ from the 
expected one (Figure~\ref{fig:2D_s3xi0}). 

It would be interesting to extend the 2D measurements to critical points statistics (peaks, voids saddles), following \citep{Bernardeau:2024ysi}. It would also be of interest to extend the present measurements  to three-dimensional flights, though we anticipate that the effect of finite boxes will become more dramatic.

Our results highlight the versatility of Rayleigh-L\'evy flights, offering insights not only for astrophysical contexts but also for broader interdisciplinary applications.
Rayleigh-Lévy flights model is a unique example of a random process that can both easily be implemented and whose  statistical properties are explicitly known (in terms of correlation functions). Furthermore those correlation functions share qualitatively the same properties as those expected for cosmic density fields in both its quasi linear regime and its strongly non linear regime.
This is therefore is very precious toy model for which analytical methods or numerical tools can be tested, from departure from Gaussian statistics to the exploration of the fully nonlinear regime.
For example, one can obtain insights into the validity range of the Edgeworth expansion, which is commonly used to model the perturbative non-Gaussianity of the matter density of the Universe as traced by galaxies. The Edgeworth expansion is not a convergent series, so understanding what variances and density ranges we can use it to model the cosmological density field is an open question. 
The paper illustrates here that fully nonlinear objects such as peaks, critical points, and their correlations, can also be investigated in such a model, challenging and extending the picture that can be drawn from Gaussian fields. Furthermore writing numerical codes to evaluate the number of peaks, droughts, critical points in a discrete set of strongly correlated points can also be challenging. Having a toy model for which exact results are known is therefore very useful.

\begin{acknowledgements}
The authors thank Jose Perico Esguerra for statistical physics insights on L\'evy flights. 
RCB and SA are supported by an appointment to the JRG Program at the APCTP through the Science and Technology Promotion Fund and Lottery Fund of the Korean Government, and was also supported by the Korean Local Governments in Gyeongsangbuk-do Province and Pohang City. SA also acknowledges support from the NRF of Korea (Grant No. NRF-2022R1F1A1061590) funded by the Korean Government (MSIT).
CP is partially supported by the grant \href{https://www.secular-evolution.org}{\emph{SEGAL}} ANR-19-CE31-0017
of the French Agence Nationale de la Recherche.
The codes underlying this work are
distributed at \href{https://github.com/reggiebernardo/notebooks/tree/main/supp_ntbks_arxiv_25xx.yyyyy}{https://github.com/reggiebernardo/notebooks}.
\end{acknowledgements}

%
%

\begin{appendix}
\onecolumn

\section{Beyond Mean Field (BMF) Calculations in 2D Rayleigh-L\'evy Flights}
\label{sec:bmf}
We present here the basis for the computation of quantities in Beyond Mean Field computation for the 2D case.
As for the 1D case described in \cite{Bernardeau:2024ysi}, the idea is to project Eq.~\eqref{treesumtau} 
on a discrete basis. 
For a Gaussian window function, the eigenfunctions of the 2D harmonic oscillator  are  a natural choice.

\subsection{The field expansion and equations}
\label{sec:fe_bmf}
Following the key  equation~\eqref{eq:defXi}, we need
to compute $\Xi_{nn'n_c}^{mm'm_c}$ for the cases of interest. In practice, for the statistics of the density PDF and $W_1$ we only need to know it for $n_c=m_c=0$ and $n_c=0$, $m_c=\pm 1$ to have access both the density and its gradient. 
In the following we will give these quantities with its explicit dependence with the sign of $m$, $m'$ and $m_c$ that will be denoted $\eps$, $\eps'$ and $\eps_c$, so that below the notations $m$ and $m'$ should  be understood as  absolute values.
Expressing the two-point function in terms of the power spectrum $P(k)$, defined such that 
$\xi(\vx)=\int\dd^2\vk\,P(k)\exp(\ii\vk.\vx)$
we have
\begin{equation}
    \Xi_{nn'0}^{\eps m\,\eps' m'0}=\int\dd^2\vk\,P(k)\,B_n^{\eps m}(\vk)\,B_{n'}^{\eps' m'*}(\vk)\,,
\end{equation}
with 
\begin{subequations}
\begin{align}
    B_n^{\eps m}(\vk)&=\int\dd^2\vx\,e^{\ii \vk.\vx}\mC_0(\vx)
    {b_n^{(m)}({\vx})}=c_n^{m}
    \int \dd r\,r^{1+{m}} \mC_0(r)
    {L_n^{(m)}(r^2/2)} \int \dd\theta \ e^{\ii \eps m\,\theta}\,e^{\ii k r \cos(\theta-\theta_k)}\,,\\
     B_n^{\eps m*}(\vk)&=\int\dd^2\vx\,e^{-\ii \vk.\vx}\mC_0(\vx)
    {b_n^{(m)}({\vx})}\,,
\end{align}
\end{subequations}
and 
\begin{equation}
    \Xi_{nn'0}^{\eps m\,\eps' m'\eps_c}=\int\dd^2\vk\,P(k)\,B_n^{\eps m}(\vk)\,B_{n'}^{\eps' m'\eps_c}(\vk)\,,
\end{equation}
    with
\begin{equation}
    B_{n}^{\eps m\,\eps_c}(\vk)=\int\dd^2\vx\,e^{-\ii \vk.\vx}\mC_1^{(\eps_c)}(\vx)
    {b_n^{(m')}({\vx})}=c_n^{m}
    \int \dd r\,r^{2+{m}} \mC_0(r)
    {L_n^{(m)}(r^2/2)} \int \dd\theta \ e^{\ii (\eps m+\eps_c)\theta}\,e^{-\ii k r\cos(\theta-\theta_k)}\,.
\end{equation}
The expressions of the normalisation coefficients $c_n^m$ can easily be obtained from \eqref{eq:orthodef}  and they read
\begin{equation}
    c_n^m=1/{\sqrt{\frac{2^m (m+n)!}{n!}}}.
\end{equation}
The integration over the angle $\theta$ yields a Bessel function of order $m$, and the final result reads
\begin{subequations}
\begin{align}
B_n^{\eps m}(\vk)&=c_n^{m}\frac{e^{-\frac{k^2}{2}} k^{m+2 n}}{(2 n)\text{!!}}\ii^m e^{\ii \eps m \theta_k}\,,
\\
B_n^{\eps m*}(\vk)&=c_n^{m}\frac{e^{-\frac{k^2}{2}} k^{m+2 n}}{(2 n)\text{!!}}\ii^{-m} e^{\ii \eps m \theta_k}\,,
\\
B_n^{\eps m\,\eps_c}(\vk)&=c_n^{m}\frac{e^{-\frac{k^2}{2}} k^{m+2 n}(-2n+(\eps \eps_c-1) m+k^2)}{(2 n)\text{!!}}\ii^{-1-m} e^{-\ii (\eps m+\eps_c) \theta_k}\,.
\end{align}
\end{subequations}
We can also express the result for a power law correlation function, or equivalently while assuming that the spectrum follows a power law in $P(k)=A\,k^{n_s}$.

For the function $W_1(\rho)$, we need to build the CGF of the density and its gradient along a given direction, say $Ox$. The CGF
will then be symmetric with respect to the $Oy$ axis. Then it suffices to restrict our basis to symmetric functions, that is to use 
\begin{equation}
    b_n^{(m)}(\vx)={\sqrt{2}}\,c_n^m\,L_n^{(m)}(r^2/2)\,r^m\,\cos(m\theta)\,,
\end{equation}
and 
$
 \mC_1(\vx)=\mC_0(r)\,r\,\cos(\theta).
$
With this definition of functions, 
Eq. \eqref{eq:defXi}
becomes
\begin{subequations}
    \begin{align}
    \Xi_{nn'0}^{mm'0}&=\xi_0\ \frac{\Gamma \left(1+{n_s}/{2}+n+n'+
   \left(m+m'\right)/2\right)}{(2 n)\text{!!} \ \sqrt{\frac{2^m (m+n)!}{n!}} \left(2
   n'\right)\text{!!} \ \sqrt{\frac{2^{m'}
   \left(m'+n'\right)!}{n'!}}\ \Gamma \left(\frac{n_s}{2}+1\right)}\delta_{m}^{m'}\,,\\
   \Xi_{nn'0}^{mm'1}&=\xi_0\,c_{m+m'}\,
   \frac{\left(\left(m\!-\!m'\right) \left(n_s\!+\!2 n\!-\! 2 n'\right)+\left(m'\!+\!m\right)\!+\!1\right)
   \Gamma \left({n_s}/{2}\!+\!n\!+n'+\!
   \left(m\!+\! m'\!+1\right)/2\right)}{4 (2 n)\text{!!}\,
   \sqrt{\frac{2^m (m+n)!}{n!}} \left(2 n'\right)\text{!!}\,
   \sqrt{\frac{2^{m'} \left(m'\!+\!n'\right)!}{n'!}}\Gamma \left(\frac{n_s}{2}+1\right)}
   \delta^{1}_{\!\abs{m'\!-\!m}}\,,
\end{align}
\end{subequations}
with $\xi_0=\Xi_{000}^{000}$, and
\begin{equation}
    c_{m+m'}=\text{if}\left[
    \begin{array}{cc}
         m+m'=1,  & \sqrt{2} \\
         m+m'>1,  & 1
    \end{array}\right].
\end{equation}
Numerical solutions were used to fix the normalization. These results can be checked by direct numerical integrations of the weighted correlation function.

\subsection{BMF density CGFs and Skewness}
\label{sec:cgfs_bmf}

With the help of the previous result, we 
can easily derive the resulting cumulant generating functions.
It only requires to solve the system \eqref{eq:fulldiscretetau}
for a chosen order. We choose here to truncate the expansion for a fixed $n_t$. For the CGF of the density only, it means that BMF results change for even indices. For the CGF for both the density and density gradient, it changes at all order.

Some results are presented in the main text. We complement here results obtained for higher orders.
For $\alpha=3/2$, the one-point density CGF with third-order polynomial corrections beyond mean field [symbolically, $\tau(\vx) \equiv \tau_0^0 + \tau_1^0 b_1^{(0)}(\vx) + \tau_2^0 b_2^{(0)}(\vx) + \tau_3^0 b_3^{(0)}(\vx)$] is given by
\begin{align}
   \varphi_6(\lambda) & = -\frac{\lambda  \left(14625 \lambda ^3 \xi _0^3-37299300 \lambda ^2 \xi _0^2+4671569920 \lambda  \xi _0-68719476736\right)}{4800 \lambda ^4 \xi _0^4-14061665 \lambda ^3 \xi _0^3+2074748004 \lambda ^2 \xi _0^2-39031308288 \lambda  \xi _0+68719476736} \,.
\end{align}
For an arbitrary power law correlation function $\xi(r) = 1/r^\beta$ the CGF $\varphi_4(\lambda)$ is given by
\begin{align}
    \varphi_4(\lambda) & \!=\! \frac{\lambda  (\beta  (\beta +2) \lambda  \xi_0  (16 (\beta  (\beta +10)+88)-\beta  (\beta +2) (\beta +4) \lambda  \xi_0 )-32768)}{\lambda  \xi_0  \left(\beta  \left(4 \beta  (\beta \!+\!2) \lambda ^2 \xi_0 ^2\!-\!(\beta  (\beta  (\beta  (\beta \!+\!8)\!+\!84)\!+\!336)\!+\!1408) \lambda  \xi_0 \!+\!16 (\beta \!+\!2) (\beta  (\beta \!+\!10)\!+\!88)\right)\!+\!16384\right)\!-\!32768}. \notag
\end{align}
CGFs with higher order BMF corrections can be obtained, but are unwieldy to print. Nonetheless, these can be calculated and stored using algebraic computing softwares such as \texttt{Mathematica}. For this work, we have evaluated up to ${\varphi}_{6}(\lambda)$ for fixed-$\alpha$ cases (e.g., $\alpha=1/2, 3/2$) and up to ${\varphi}_{4}(\lambda)$ for the arbitrary power law correlation.

\begin{figure}[h!]
    \centering
    \includegraphics[width=0.99\linewidth]{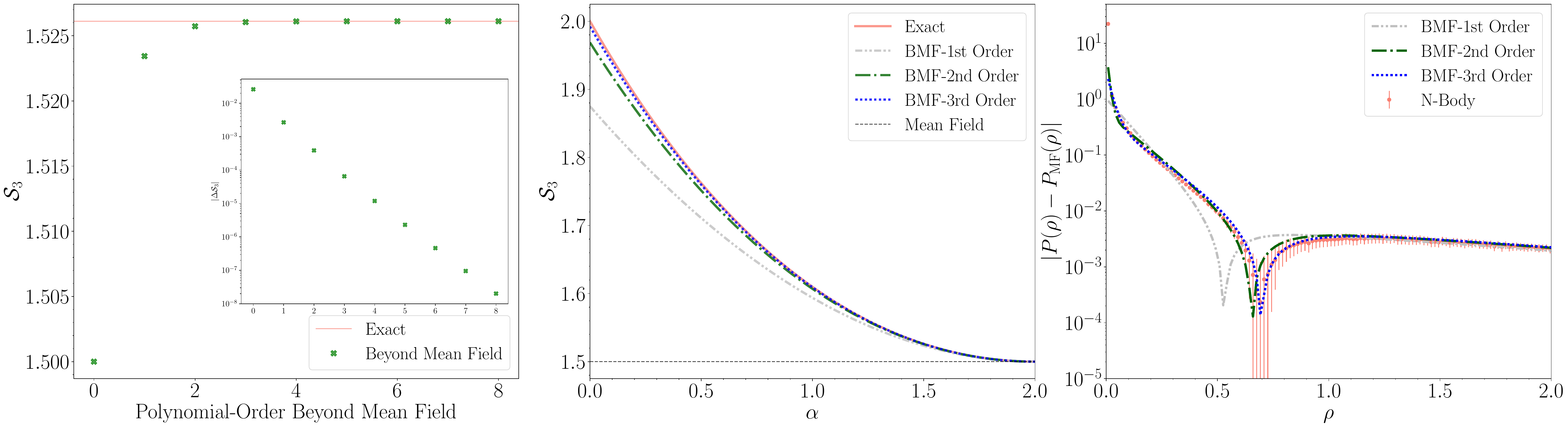}   
    \caption{Skewness in two-dimensional L\'evy flight distribution in the `beyond mean field' formalism ({\sl left}) as a function of the order of the expansion in \eqref{eq:r_cgf_expansion} for $\alpha=3/2$ and ({\sl middle}) as a function of the L\'evy flight index $\alpha$. Exact and mean field (MF) values are shown as {red-solid} and black-dotted lines, respectively. ({\sl Right}) The difference between mean field and beyond mean field density probability distribution functions for different expansion orders. We present the PDF from the main body of the text; $\xi_{0} = 2.47$ and use an effective $\alpha =1.43$ in the BMF expansion. The red points/error bars are the mean and error on mean of the simulated flights.}
    \label{fig:S3_2D_BMF}
\end{figure}

The skewness obtained for CGFs with increasing order of corrections beyond mean field are shown in Figure \ref{fig:S3_2D_BMF}, supporting an exponential convergence of the BMF formalism to the exact solution.

\subsection{BMF joint density and density gradient CGF}

The first non trivial result is obtained for $\alpha=3/2$ and when we restrict to $n_t=1$, that is when the first two functions of the basis only are used, is given by 
\begin{equation}
\varphi_1(\lambda,\lambda_x)=\frac{16 \left(16 \lambda +\xi _0 \left(\lambda _x^2-\lambda ^2\right)\right)}{\xi _0^2 \left(8 \lambda ^2-7 \lambda _x^2\right)+256 -144 \lambda   \xi _0}\,.   \label{eq:phi1}
\end{equation}
Its counterpart  for any correlation index $\beta$ is given by Eq.~\eqref{eq:phi1MT} in the  main text, taking advantage of the fact that all correlators are explicitly known.
The next interesting order is $n_t=3$. At such order we can include the next basis element for $m=1$. Note that to be consistent, there are in total six basis elements to include, 
$n=0$ and $1$ for $m=0$ and $1$ and $n=0$ for $m=2$ and $m=3$. 
For $\alpha=3/2$, the result takes the form
\begin{align}
\varphi_3(\lambda,\lambda_x)&=\frac{P(\lambda,\lambda_x)}{Q(\lambda,\lambda_x)}\,, \quad \mathrm{where}
\label{eq:phi3}\\
P(\lambda,\lambda_x)&=
4 \left(7.03687\times 10^{13} \lambda +2.51658\times 10^8 \lambda  \xi _0^2 \left(2061
   \lambda ^2-1880 \lambda _x^2\right)\right.\nonumber\\
   &\left. -1.37439\times 10^{11} \xi _0 \left(77 \lambda ^2-32
   \lambda _x^2\right)+500 \lambda  \xi _0^4 \left(168832 \lambda ^4-522016 \lambda ^2
   \lambda _x^2+183207 \lambda _x^4\right)\right.\nonumber\\
   &\left.-204800 \xi _0^3 \left(50344 \lambda ^4-87243.
   \lambda ^2 \lambda _x^2+9792 \lambda _x^4\right)-1875 \xi _0^5 \left(128 \lambda ^6-608
   \lambda ^4 \lambda _x^2+513 \lambda ^2 \lambda _x^4-105 \lambda
   _x^6\right)\right)\,,\nonumber\\
   Q(\lambda,\lambda_x)&= 6.71089\times 10^7
   \xi _0^2 \left(329923 \lambda ^2-123944 \lambda _x^2\right)\nonumber\\
  & -1.83069\times 10^{14} \lambda  \xi _0-32768 \lambda  \xi _0^3
   \left(2.918\times 10^7 \lambda ^2-2.57737\times 10^7 \lambda _x^2\right)\nonumber\\
   &-100 \lambda  \xi _0^5
   \left(1.38483\times 10^6 \lambda ^4-4.16963\times 10^6 \lambda ^2 \lambda
   _x^2+1.45172\times 10^6 \lambda _x^4\right)\nonumber\\
   &+240 \xi _0^4 \left(7.34829\times 10^7 \lambda
   ^4-1.24732\times 10^8 \lambda ^2 \lambda _x^2+1.41909\times 10^7 \lambda _x^4\right)\nonumber\\
   &+375
   \xi _0^6 \left(1024 \lambda ^6-4736 \lambda ^4 \lambda _x^2+3912 \lambda ^2 \lambda
   _x^4-795 \lambda _x^6\right)+2.81475\times 10^{14}\nonumber \,.
\end{align}
This is the form used in practice to compare with simulations in the computation of the $W_1$ indicator.

\section{Derivation of $W_1(\rho)$ as a function of the CGF}
\label{sec:app_b}
The  iso-field perimeter, $W_1$, is given in general by Eq.~\eqref{eq:defW1}
\begin{equation}
    W_1(\rho)
    =
    \iint \sqrt{\rho_x^2+\rho_y^2}\ \dd\rho_x\dd\rho_y\ P(\rho,\rho_x,\rho_y)\,,
    \label{eq:W1app}
\end{equation}
where the PDF, $P(\rho,\rho_x,\rho_y)$, can be expressed  as  
\begin{equation}
P(\rho,\rho_x,\rho_y)=
\int\frac{\dd\lambda_\rho}{2\pi\ii}
\int\frac{\dd\lambda_x}{2\pi\ii}
\int\frac{\dd\lambda_y}{2\pi\ii}
\exp\left(
-\lambda_\rho\rho-\lambda_x\rho_x-\lambda_y\rho_y+
\varphi(\lambda,\lambda_x,\lambda_y)
\right)\,, \label{eq:defPDFapp}
\end{equation}
in terms of  $\varphi(\lambda,\lambda_x,\lambda_y)$,  the joint cumulant generating function for $\rho$, $\rho_x$ and $\rho_y$. 
After integration by parts, Eq.~\eqref{eq:defPDFapp} becomes
\begin{equation}
P(\rho,\rho_x,\rho_y)=
\int\frac{\dd\lambda_\rho}{2\pi\ii}
\int\frac{\dd\lambda_x}{2\pi\ii}
\int\frac{\dd\lambda_y}{2\pi\ii}
\frac{1}{\rho_x}
\frac{\partial\varphi}{\partial\lambda_x}
\exp\left(
-\lambda_\rho\rho-\lambda_x\rho_x-\lambda_y\rho_y+
\varphi(\lambda,\lambda_x,\lambda_y)
\right) \,.
\end{equation}
The integration element $\dd \rho_x \dd \rho_y$ and $\dd\lambda_x\dd\lambda_y$ in Eqs.~\eqref{eq:W1app} and~\eqref{eq:defPDFapp} can be written respectively
in polar coordinates as 
$\rho_g\dd\rho_g\dd\theta_g$ and $\lambda_g\dd\lambda_g\dd\theta_\lambda$. We further take advantage of the fact that 
$\varphi(\lambda,\lambda_x,\lambda_y)$ is a function of $\lambda_g^2\equiv
\lambda_x^2+\lambda_y^2$ only so that
\begin{equation}
    \frac{\partial\varphi}{\partial\lambda_x}=
    \frac{\lambda_x}{\lambda_g}
    \frac{\partial\varphi}{\partial\lambda_g}\,.
\end{equation}
This implies that Eq.~\eqref{eq:W1app} can we re-expressed explicitly in terms of the cumulant generating function  as
\begin{equation}
    W_1\!=\!\!\!
    \iiint \frac{\dd\lambda_\rho}{2\pi\ii}
    \frac{\lambda_g\dd\lambda_g\dd \theta_\lambda}{2\pi^2}
   \!\! \iint \rho_g^2 \dd \rho_g\dd \theta_g
\frac{1}{\rho_g \cos\theta_g}\frac{\lambda_g\cos\theta_\lambda}{\lambda_g}
\frac{\partial\varphi(\lambda_\rho,\lambda_g)}{\partial\lambda_g}
\exp\left(
-\lambda_\rho\rho-\lambda_g\rho_g\cos(\theta_\lambda-\theta_g)+
\varphi(\lambda,\lambda_g)
\right)\,. \label{eq:W1app2}
\end{equation}
The integral over $\theta_\lambda$ in Eq.~\eqref{eq:W1app2} gives
\begin{equation}
    \int \dd \theta_\lambda
    \cos\theta_\lambda
    \exp\left(\lambda_g\rho_g\cos(\theta_\lambda-\theta_g)\right)
    =-2\pi \ii J_1(\lambda_g\rho_g)\cos\theta_g \,.
\end{equation}
Using then the fact that
\begin{equation}
    J_1(\lambda_g\rho_g)=-\frac{1}{\rho_g}\frac{\dd}{\dd\lambda_g}
     J_0(\lambda_g\rho_g)\,,
     \quad \mathrm{and} \quad
    \int_0^\infty\dd\rho_g J_0(\lambda_g\rho_g)=\frac{1}{\lambda_g}\,,
\end{equation}
allows us to carry out the
integration over $\rho_g$ and $\theta_g$, so that $W_1$ becomes
\begin{equation}
    W_1=-
   \int_{-\ii\infty}^{+\ii\infty} \frac{\dd\lambda_\rho}{2\pi}
    \int_0^\infty\frac{\dd\lambda_g}{\lambda_g}
\frac{\partial\varphi(\lambda_\rho,\lambda_g)}{\partial\lambda_g}
\exp\left(
-\lambda_\rho\rho+
\varphi(\lambda_\rho,\lambda_g)
\right)    \,,
\end{equation}
which in practice can be rewritten as Eq.~\eqref{eq:W1final} (given in the main text), after a further integration by parts.

\section{Numerical Algorithms and Testing}
\label{sec:numerical_tests}

In this Section we briefly describe the numerical algorithms used to generate the density field PDF and Minkowski Functional curves. L\'evy flights are generated sequentially as point distributions that are binned onto a regular lattice using a nearest neighbour scheme. Throughout this work we take a regularly spaced lattice with $N=4000$ points per side and resolution $\Delta = 0.25$, generating a box length $L = 4000 \times \Delta=1000$. Periodic boundary conditions are enforced modulo $L$ at each step of the flight, imposing periodic boundary conditions on the resulting field. 

The density field constructed by binning the flight is subjected to a discrete Fourier transform, then smoothed in Fourier space using a kernel $W(kR) = \exp[-k^{2}R^{2}/2]$ and inverse Fourier transformed. The smoothing scale is always set to be $R=1$, to match the Gaussian filter used in the theoretical calculations. From these fields we extract the summary statistics presented in the main body of the text. The PDF of the field is estimated by binning the pixels into a $\rho_{i}$ array, and then normalising the resulting histogram to unity. We take $N=500$, $\rho_{i}$ values regularly spaced over $0 < \rho \leq 10$, except for the case $\xi_{0} = 27$ in which case we use $N=1200$ values regularly spaced over $0 < \rho \leq 80$. 

The Minkowski Functionals $W_{1}$ and $\chi$ are extracted from the field by taking the array of density threshold values $\rho_{i}$, and generating iso-field boundaries for each threshold. A constant field boundary is a one-dimensional curve, which is generated using the marching squares algorithm as detailed in \cite{Appleby:2017uvb}. In short, the algorithm linearly interpolates between adjacent values of the density field on the lattice according to a scheme that creates a set of closed polygons. $W_{1}$ is then the total length of the line segments that make up the polygons, and we define the genus $W_{2}$ as the sum of rotation angles along the iso-field lines, divided by $2\pi$. The Euler characteristic $\chi$ is then the derivative of $W_{2}$ with respect to $\rho_{i}$; we use the central derivative scheme $\chi_{i} = (W_{2,i+1}-W_{2,i-1})/(\rho_{i+1}-\rho_{i-1})$. Both $W_{1}$ and $\chi$ are defined here as densities, so they are normalized by the area of the box. 

\subsection{Gaussian Random Fields}

To assess the accuracy of the algorithms adopted, we extract the Minkowski Functionals from fields for which exact expectation values are known. We cannot use L\'evy flights with large $\xi_{0}$ variances to test the numerics in this way because their expectation values are not exactly known; they are subject to BMF corrections and the convergence of this approximation to the exact properties of the field is unknown. In this appendix we generate Gaussian random fields, for which the ensemble expectation values for $P(\rho)$, $W_{1}$ and $\chi$ are exactly known. This allows us to quantify the numerical limits of our methodology, at least for fields that are close to Gaussian. 

\begin{figure}[h!]
    \centering
    \includegraphics[width=0.98\linewidth]{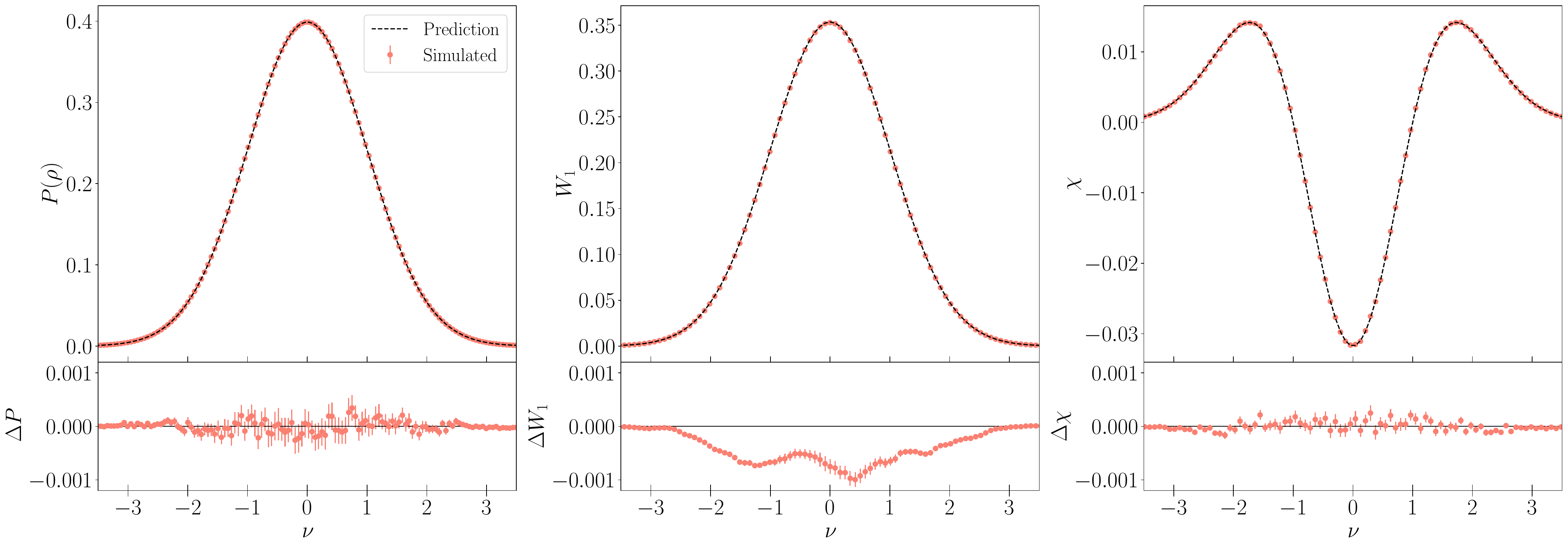}\\
    \caption{Summary statistics (PDF, $W_{1}, \chi$) extracted from Gaussian random fields ({\sl left, middle, right panels}). The red points/errorbars are measured from simulations and the black dashed lines are the expectation values. The lower sub-panels show the difference between measurements and theory.  }
    \label{fig:2D_num_check}
\end{figure}

In Figure \ref{fig:2D_num_check} we present the summary statistics $P, W_{1}, \chi$ extracted from $N_{\rm real} = 50$ realisations of a Gaussian random field. Red points/errorbars are the sample means and error on the means of the realisations. Specifically, the Gaussian random fields are drawn in Fourier space from a white noise power spectrum $P(k) \propto k^{n}e^{-k^{2}R^{2}}$ with $n=0$ and the same parameters as the main body of the text; $R = 1$, $\Delta = 0.25$, $L=4000\times \Delta$. After inverse fast Fourier transforming, the density field is mean subtracted and normalised to unit variance; $\nu = (\rho - \bar{\rho})/\sqrt{\xi_{0}}$. For Gaussian fields, the PDF and Minkowski Functionals have the exact ensemble expectation values 
\begin{equation}  P(\nu) = {1 \over \sqrt{2\pi}}e^{-\nu^{2}/2}\,, \quad
 W_{1}(\nu) = {1 \over 2} \sqrt{\xi_{1} \over 2\xi_{0}}e^{-\nu^{2}/2} \, ,\quad
 \chi(\nu) = {1 \over (2\pi)^{3/2}}{\xi_{1} \over 2 \xi_{0}} e^{-\nu^{2}/2}\left( \nu^{2}-1 \right) \,.
\end{equation} 
We compare the numerically extracted statistics (red points/error bars) to the Gaussian expectation (black dashed lines). The sub-panels show $\Delta P$, $\Delta W_{1}$ and $\Delta \chi$; the differences between theoretical predictions and simulated measurements for each statistic (red points/error bars). For each statistic, the simulation measurements successfully reproduce the predictions to sub-percent level. $W_{1}$ is the only statistic that exhibits any systematic offset (cf middle panel); this is because the linear interpolation approximation of a smooth curve neglects curvature at the resolution scale. Even for $W_{1}$, the systematic offset between numerical reconstruction and prediction $\lesssim 0.5\%$ for all densities.

\end{appendix}

\end{document}